\documentstyle[preprint,eqsecnum,aps]{revtex}

\begin{document}
\draft

\widetext

\title{Class of colliding plane waves in terms of Jacobi functions}

\author{
Nora Bret\'on$^{\star}$\thanks{E-mail: nora@fis.cinvestav.mx},
Alberto Garc\'{\i}a$^{\star \P}$\thanks{E-mail:  
agarciad@lauca.usach.cl}, 
Alfredo Mac\'{\i}as$^\diamond$\thanks{E-mail: amac@xanum.uam.mx},
and Gustavo Y\'a\~nez$^{\star}$\thanks{E-mail: gyanez@fis.cinvestav.mx}\\
$^{\star}$  Departamento de F\'{\i}sica,\\
CINVESTAV--IPN, Apartado Postal 14--740, C.P. 07000, M\'exico,  
D.F., MEXICO\\
$^{\P}$  Departamento de F\'{\i}sica,\\
Universidad de Santiago de Chile,\\
Avda. Ecuador 3493--Casilla 307.Correo 2, Santiago,  CHILE\\
$^{\diamond}$ Departamento de F\'{\i}sica,\\
Universidad Aut\'onoma Metropolitana--Iztapalapa,\\
Apartado Postal 55--534, C.P. 09340, M\'exico, D.F., MEXICO.\\}

\date{\today}

\maketitle

\begin{abstract}
We present a general class of noncolinear colliding wave solutions of the 
Einstein--Maxwell equations given in terms of fourth order polynomials,
which in turn can be expressed through Jacobi functions depending on 
generalized advanced and retarded time coordinates. The solutions are
characterized by six free parameters. The parameters can be chosen in such a
way to avoid the generic focusing singularity.
\end{abstract}
\vspace{0.5cm}


\narrowtext
\section{Introduction}
Recently, the collision of plane--fronted gravitational waves
possibly coupled with electromagnetic waves, has been extensively
studied \cite{khan,szek,chandra1,chandra2}. Because gravity is always
attractive, it was expected that focusing of the waves would occur and one of 
the interesting questions is how much focusing does general relativity predict.
Within this framework, strong focusing would appear by the development of
spacetime curvature singularities.
Many solutions has been presented so far, describing  collisions of 
plane--fronted gravitational and electromagnetic waves. And quite a few of 
them do develop Cauchy horizons. It is important to stress the fact that
all of these solutions contain {\em only second degree} polynomials and
that till now nobody has gone ahead to {\em higher degree polynomials} in 
the sense we are going to explain below.

All known colliding wave solutions are characterized, in null coordinates, 
by the line element
\begin{equation}
ds^2= e^{-M}du dv+g_{ab} dx^a dx^b, \quad a,b=1,2, 
\end{equation}
where $x^1$, $x^2$ are ignorable coordinates and  metric functions 
depending on $(u,v)$ \cite{grifiths}. When written in $(\eta, \mu)$ 
coordinates \cite{chandra1}, where $\eta$ is a measure of the  time from the 
instant of collision and $\mu$ is a measure of the distance between the plane 
wavefronts, the metric becomes
\begin{equation}
ds^2= e^{2 \gamma}(\eta^2-\mu^2) \left( {d\eta^2 \over P(\eta)} 
- {d\mu^2 \over Q(\mu)}\right) + g_{ab} dx^a dx^b,
\end{equation}
where the polynomials $P(\eta)$, $Q(\mu)$ are at most of the second degree in 
their respective variables.
Generating techniques and other methods have allowed a great deal of 
generalizations in relation with the metric components $g_{ab}$ 
(sometimes called Killingian sector) \cite{e1,e3,ernst,e4,f1,f2,f3,f4}. The 
generated solutions preserve the seed's structure of the polynomials $P(\eta),
Q(\mu)$, changing in general the form of the metric function $\gamma$.

In this paper, we pursue further and present solutions with $P(\eta)$ and 
$Q(\mu)$ being polynomials of fourth degree, by means of a generalization of 
the advanced and retarded coordinates concept. This generalization endows the 
corresponding spacetimes with a geometric structure for the time and 
independently for the 
spatial coordinate richer than the already known colliding wave solutions.
In the limit of polynomials of second degree one recovers the widely used 
retarded and advanced time coordinates $(u,v)$.
The main  purporse of this paper is to give the formulation of the most 
general class nowadays known of colliding waves with fourth degree polynomials. 
As usual, it is assumed that in the corresponding spacetime, the two 
waves approach each other from opposite sides in flat Minkowski background; 
after the collision, a new gravitational field evolves, which satisfies 
certain continuity conditions. The 
colliding plane waves possess five symmetries, while the geometry resulting 
after the collision has two spacelike Killing vectors.    

The plan of the paper is as follows: In Sec. II we revisited some generalities
on colliding waves concept. In Sec. III the general representation 
through advanced and retarded time coordinates for fourth degree polynomials 
is introduced. In Sec. IV the class of colliding waves with fourth degree 
polynomials is presented. In Sec. V the results are discussed.

\section{Colliding waves}

As was stated previously, in this section we review briefly some 
generalities on electrovacuum colliding waves, introduced by Ernst et al. 
\cite{ernst}.
 
The set of colliding waves solutions is
described by the line element 
\begin{equation} 
g=2g(u,v) du dv + g_{ab}(u,v) dx^a dx^b \, , \qquad\, a,b=1,2\, , 
\end{equation}  
where $x^1 = x$, $x^2=y$ are ignorable coordinates.  The domain of the
coordinate charts consists of $(x,y) \in {\bf R}^2$ and $(u,v)\in
{\bf R}^2$; it is the union of four continuous regions: $I:=
\{(u,v): 0\leq u < 1, 0\leq v < 1\}$ , $II:= \{(u,v): u\leq 0,
 0\leq v < 1\}$, $III:= \{(u,v): 0\leq u < 1, v < 0\}$,
$IV:=\{(u,v): u\leq 0, v\leq 0\}$. 

In the region $IV$, a closed subregion of the Minkowski space, it  
is required that
\begin{equation} 
g_{\mu \nu}(u,v)= g_{\mu \nu}(0,0)\, , \quad\,  A_{\mu}(u,v)= A_\mu(0,0)\, , 
\end{equation} 
where $A_{\mu}$ is the electromagnetic vector potential.
By scaling--shifting transformations the metric can be brought to  
standard Minkowski metric.
In region $II$, the metric components  and the electromagnetic vector 
potential
depend only on $v$ i.e. $g_{\mu \nu}=g_{\mu \nu}(0,v),\, 
A_{\mu}=A_{\mu}(0,v)$. 
In region $III$ these fields are 
functions of the coordinate $u$, i.e. $g_{\mu\nu}=g_{\mu\nu}(u,0),  
A_{\mu}=A_{\mu}(u,0)$. In region $I$, which is  
occupied by the scattered null fields, the metric components and the 
electromagnetic field  are functions of both $u$ and $v$ coordinates.        
In this way, the problem is reduced to know the metric (coframe) and the 
electromagnetic vector fuction $A_{\mu}$.

\subsection{Newman-Penrose tetrad}

We introduce a complex null frame according to the conventional
description: 
\begin{equation}
{e}_\alpha=({\bf l},{\bf n},{\bf m},{\bf \bar{m}}) 
\label{nullframe} \,
\end{equation}
with 
\begin{equation}
{\bf l}^2={\bf n}^2={\bf m}^2={\bf\bar{m}}^2 ={\bf l}\cdot{\bf m}={\bf n}
\cdot{\bf m}={\bf l}\cdot{\bf\bar{m}}={\bf n}\cdot{\bf\bar{m}}=0
\label{nullo} \, ,
\end{equation} 
\begin{equation}
{\bf l}\cdot{\bf n}=1,\quad{\bf m}\cdot{\bf\bar{m}}=-1
\label{one} \, .
\end{equation}
The basis of the corresponding
$1$--forms, the coframe, is denoted by $\vartheta^\alpha$. If we lower
the coframe index, then we use the same letters as with the frame:
$\vartheta_\alpha=({\bf l},{\bf n},{\bf m},{\bf \bar{m}})$. Then
the coframe reads  
\begin{equation}
\vartheta^\alpha=({\bf n},{\bf l}, -{\bf\bar{m}},-{\bf {m}})
\label{coframe1}\, .
\end{equation}
According to (\ref{nullo}) and (\ref{one}), we have 
\begin{equation}
g=2\left(\vartheta^{\hat{0}}\otimes\vartheta^{\hat{1}}-
\vartheta^{\hat{2}} \otimes\vartheta^{\hat{3}}
\right)=n_{\alpha\beta}\,\vartheta^\alpha\otimes\vartheta^\beta
\label{metric}\, ,
\end{equation}
with 
\begin{equation}
n_{\alpha\beta} =
\left(
  \begin{array}{cccc}
    0 & 1 & 0 & 0\\
    1 & 0 & 0 & 0\\
    0 & 0 & 0 &-1\\
    0 & 0 &-1 & 0
  \end{array}
\right)
\label{nullmetric}\, .
\end{equation}

\subsection{Self--dual part of the complex Weyl $2$--form}

Quite independently from the type of coframe we are using, we can
consider some properties of the Weyl 2--form. Let us denote the Weyl
curvature 2--form by \begin{equation}\label{Weyl}
C_{\alpha\beta}=\frac{1}{2}\,C_{\mu\nu\alpha\beta}\,\vartheta^
\mu\wedge\vartheta^\nu\,.  \end{equation} 

If, in four dimensions, an arbitrary complex $2$--form $\omega$ is given,
then its Hodge dual $^\star\omega$ is again a 2-form. Therefore we can
build linear combinations of both: \begin{equation}\label{combination}
\omega^{\pm}:=\frac{1}{2}\left(\omega\pm i\,^\star\omega\right)\,.
\end{equation} Here $i$ is the imaginary unit. We call $\omega^+$ the
self--dual and $\omega^-$ the antiself--dual part of the $2$--form
$\omega$. Clearly, 
\begin{equation}
\omega=\omega^++\omega^-
\label{sum}\, .
\end{equation} 
A form $\omega$ is called self--dual, if its
antiself--dual piece vanishes, i.e., if $\omega=i\,^\star\omega$, it is
called antiself--dual, if $\omega=-i\,^\star\omega$.

Let us now assume that the form $\omega$ is {\em real}, that is,
$\omega=\overline{\omega}$, where the bar denotes complex
conjugation. Then, according to (\ref{combination}), they fulfill the
relations: 
\begin{equation}
\overline{\omega^+}=\omega^-\,,\qquad{\rm and}\qquad
\overline{\omega^-}=\omega^+ 
\label{dual}\, .
\end{equation} 
Consequently, a real
$2$--form $\omega=\overline{\omega}$ can be alternatively encoded into
its self--dual part $\omega^+$. Incidentally, the same is true for its
antiself--dual part.

The Weyl $2$--form is a real quantity. Accordingly, we can also take its
self--dual part 
\begin{equation}
{C}^+_{\alpha\beta}:=\frac{1}{2}\left({C}_{\alpha\beta}
+i\,^\star{C}_{\alpha\beta} \right)
\label{selfdual}\, .
\end{equation} 
Expanding the Weyl $2$--form ${C}^+_{\alpha\beta}$ in terms of the basis
$\vartheta^\mu\wedge\vartheta^\nu$ we have 
\begin{equation}
{C}^+_{\alpha\beta}=\frac{1}{2}\,{C}^+_{\mu\nu\alpha\beta}
\,\vartheta^\mu\wedge\vartheta^\nu
\label{expansion}\, ,
\end{equation}
which we will asumme to be in the standard $2$--form selfdual basis 
$[{\bf U}, {\bf V}, {\bf W}]$. The most general form of our Weyl field 
components, compatible with  
colliding wave spacetime structure is given by \cite{alberto}:
\begin{eqnarray}
A_{\mu}&=&A_{\mu}(u,v)  , \, \,  
C^+_{abcd} = 2\Psi_0 U_{ab} U_{cd} + 2\Psi_2 (U_{ab} V_{cd} 
+ V_{ab} U_{cd}+ W_{ab} W_{cd})+2\Psi_4 V_{ab} V_{cd}
\, , {\rm region\, I} , \nonumber \\ 
A_{\mu}&=&A_{\mu}(v) \, , \quad C^+_{abcd} = 2\Psi_0 U_{ab} U_{cd}\,  
,\quad  
{\rm region\, II}\, ,\nonumber\\
A_{\mu}&=&A_{\mu}(u)\, , \quad C^+_{abcd} = 2 \Psi_4 V_{ab}   
V_{cd}\, ,\quad
{\rm region\, III} \, ,
\end{eqnarray}
with
\begin{eqnarray}
W_{ab}&=&  m_a \tilde m_b - m_b \tilde m_a - k_a l_b+ k_bl_a \,  
,\nonumber\\ 
V_{ab}&=&  k_a  m_b - k_b  m_a \,  , \nonumber\\
U_{ab} &=& -l_a \tilde m_b +l_b \tilde m_a \, .
\end{eqnarray} 
where $m_a, \tilde m_b, k_a$ and $l_a$ are null tetrads.

\section{Generalized advanced and retarded time coordinates}

We shall consider a metric of the form
\begin{equation}
g = e^{ - 2 \gamma}  \left( \frac{dp^2}{P(p)} 
- \frac{dq^2}{Q(q)}\right) + g_{ab}dx^a dx^b,
\label{2.5}\, , 
\end{equation}
where $P, Q$ are fourth degree polynomials in their respective variable,
and $\gamma$ depending on both $(p,q)$ coordinates.

Since we have real polynomials of  fourth degree, when looking for 
advanced and retarded time variables one has to deal with elliptic integrals,
i.e. we shall use the Legendre first kind integrals:
\begin{equation}
\int \frac{dr}{\sqrt{G_4(r)}} = \mu \int \frac{d\phi}{\sqrt{1-k^2 \sin^2 \phi}}
=\mu F(\phi,k)\, , 
\end{equation}  
where the fourth degree polynomial $G_4(r)$, is 
represented by the product of four monomials containing the corresponding 
roots, the coefficient of the higher degree (fourth), denoted by $a_4$, is 
assumed to be 
$a_4 = \pm 1$; if this were no the case, i. e. $a_4 = \pm \alpha_4, \alpha_4 > 0$, 
then dividing by $\alpha_4$ one arrives at the above equation where now $ \mu$ 
has to be interpreted as $\frac{\mu}{\sqrt{\alpha_4}}$, in what follows we 
shall adopt 
these conventions when needed. The real roots are denoted by $r_j$, with 
$j=1,\dots , 4$ and $r_1 > r_2 > r_3 > r_4$, while the complex roots by 
$s_1 \pm i t_1$,
$s_2 \pm i t_2$, with $s_1 \ge s_2$, $t_1>t_2>0$, we introduce also the 
following useful notation
\begin{eqnarray}
r_{ik} &=& r_k - r_i, \, \quad (i,k=1,2,3,4)\, , \nonumber\\ 
\left(r,\beta,\gamma,\delta  \right) &=& \frac{r-\gamma}{r-\delta}\, 
\frac{\beta-\delta}{\beta-\gamma}\, , \nonumber \\
\tan \theta_1 &=& \frac{r_1- s_1}{t_1}\, , \quad \tan \theta_2 = 
\frac{r_2- s_1}{t_1}\, , \nonumber \\
\tan \theta_3 &=& \frac{t_1 + t_2}{s_1-s_2} \, , \quad
\tan \theta_4 = \frac{t_1 - t_2}{s_1-s_2}\, , \nonumber \\
\tan \left[(\theta_5/2)^2\right] &=& \frac{\cos \theta_3}{\cos \theta_4}\, ,
\nonumber \\
\nu &=& \tan \left[(\theta_2 - \theta_1)/2\right] \tan \left[(\theta_2 + 
\theta_1)/2\right]\, . 
\end{eqnarray}
The elliptic integral $F(\phi,k)$ is the Legendre integral of the first kind.
As it is well known, the standard form for writing this function is
\begin{equation}
z=\int^\phi \frac{d\phi}{\sqrt{1-k^2 \sin^2 \phi}}=F(\phi,k)\, , \quad 
\phi = {\rm am}\, z\, ,
\end{equation}
where ${\rm am}\, z$ denotes the function amplitude of $z$.
Replacing $\phi$ through $\omega$ according to 
\begin{equation}
\omega = \sin \phi= \sin ({\rm am}\, z) := {\rm sn}\, z\, , \quad  
z= \int_{0}^\omega \frac{d\omega}{\sqrt{(1-\omega^2) (1-k^2  \omega^2)}}=
{\widetilde F}(\omega,k)\, , 
\end{equation}
where ${\rm sn}\, z$ belongs to the Jacobi family of elliptic functions 
$({\rm sn}\, z, {\rm cn}\, z, {\rm dn}\, z)$, with well established analytical
properties. The values and main properties of these function can be found in 
\cite{korn}.

Going back to the line element (\ref{2.5}) and assuming that 
$P(p)$ and $Q(q)$ are polynomials up to fourth degree
on $p$ and $q$ respectively, then 
the two dimensional line element, first two terms in (\ref{2.5}), can be 
written in terms of the retarded and advanced time coordinates 
${\widetilde u}$ and ${\widetilde v}$, respectively, as follows 
\begin{equation}
e^{-2 \gamma(p,q)} \left( \frac{dp^2}{P(p)} - \frac{dq^2}{Q(q)} \right) 
= e^{- \gamma( \widetilde u, \widetilde v )}d{\widetilde u} d{\widetilde v}\, ,
\end{equation}
with 
\begin{equation}
\left. \begin{array}{l}
d{\widetilde u} \\ 
d{\widetilde v}
\end{array} \right\}  =\frac{dp}{\sqrt{P_4(p)}} \pm  \frac{dq}{\sqrt{Q_4(q)}} 
= \frac{\mu_\phi}{\sqrt{\alpha_4}}
\frac{d\phi}{\sqrt{1-k_\phi^2 \sin^2 \phi}} \pm 
\frac{\mu_\theta}{\sqrt{\beta_4}} 
\frac{d\theta}{\sqrt{1-k_\theta^2 \sin^2 \theta}} 
\label{beto1}\, ,
\end{equation}
where $\mu_{\phi}, \mu_{\theta}$ are constants and $\alpha_4, \beta_4$ are the
moduli of the coefficients of the higher power in the polynomials $P$ and $Q$,
correspondingly. The variables ${\widetilde u}$ and ${\widetilde v}$ range 
the values from $-\infty$ to $+\infty$.

The integration of (\ref{beto1}) yields 
\begin{equation}
\left. \begin{array}{l}
{\widetilde u} \\
{\widetilde v}
\end{array} \right\} = \frac{\mu_\phi}{\sqrt{\alpha_4}} F(\phi, k_\phi) \pm  
\frac{\mu_\theta}{\sqrt{\beta_4}} 
F(\theta, k_\theta) 
\label{beto2}\, ,
\end{equation}
where $-\infty < {\widetilde u} < \infty$, $-\infty < {\widetilde v}< 
\infty$. 
Hence, one has relations between the Legendre elliptic integrals and the
null coordinates $\widetilde u$, $\widetilde v$, i.e.
\begin{equation}
F(\phi, k_\phi) = \frac{\sqrt{\alpha_4}}{2\mu_\phi}   
\left({\widetilde u}+{\widetilde v}
\right)\, , \quad F(\theta, k_\theta) = \frac{\sqrt{\beta_4}}{2\mu_\theta} 
\left({\widetilde u}-{\widetilde v}
\right)\, .   
\end{equation}
Consequently, the inversion formulas bring the following functional dependence
upon the generalized advanced and retarded time coordinates
\begin{eqnarray}
\phi&=& {\rm am} \left[\frac{\sqrt{\alpha_4}}{2\mu_\phi}\left({\widetilde u}
+{\widetilde v}
\right)\right]\, , \quad \omega= \sin \phi = {\rm sn} 
\left[\frac{\sqrt{\alpha_4}}{2\mu_\phi}\left({\widetilde u}+{\widetilde v}
\right)\right]\, ,\\
\theta&=& {\rm am} \left[\frac{\sqrt{\beta_4}}{2\mu_\theta} 
\left({\widetilde u}-
{\widetilde v}\right)\right]\, , \quad \Omega= \sin \theta = {\rm sn} 
\left[\frac{\sqrt{\beta_4}}{2\mu_\theta} \left({\widetilde u}
-{\widetilde v}\right)\right]\, ,
\end{eqnarray}
which allow us to give the original $p$ and $q$ coordinates, initially 
expressed through $\phi$ and $\theta$, in terms of $\widetilde u$ and 
$\widetilde v$ null coordinates, i.e. 
\begin{equation}
p=p({\widetilde u},{\widetilde v})\, , \quad {\rm and}\quad 
q=q({\widetilde u},{\widetilde v})
\label{betito}\, .
\end{equation}

Moreover, one can introduce a new set of variables $(u,v)$ 
\begin{equation}
u= \sin \left({ \frac {\widetilde u}{2}} \right), \quad {\rm and}\quad 
v= \sin \left({ \frac {\widetilde v}{2}} \right).
\end{equation}
where the ranges of the variables are $- \infty < \widetilde u < \infty$,
$- \infty < \widetilde v < \infty$ and $-1<u<1, \quad -1<v<1$. These 
variables are such that
\begin{equation}
d{\widetilde u}= 2 \frac{du}{U}, \quad d{\widetilde v}= 2 \frac{dv}{V}
\end{equation}  
and
\begin{equation}
\frac{\widetilde u\pm \widetilde v}{2}=\arcsin u \pm \arcsin v=
\arcsin [uV \pm vU]
\end{equation}
where $U= \sqrt{1-u^2}, V= \sqrt{1-v^2}$ and the corresponding term 
in the line element  
\begin{equation}
d \widetilde u \otimes d \widetilde v = 4 \frac{du}{U} \otimes \frac{dv}{V}.
\end{equation}
Having in mind colliding wave interpretations of the results, one 
restricts $u$ and $v$ to the range $0\leq u < 1$, $0\leq v < 1$
to be closed to the standard colliding wave approach.
Depending on the character, real or complex, of the 
roots of the fourth degree polynomials, one may encounter in 
general combinations of different possible cases.
It is always possible to introduce the retarded and 
advanced time coordinates $u$ and $v$. However, only certain solutions 
satisfy the Ernst \cite{ernst} requirements for colliding waves. We shall 
restrict ourselves to metrics which can support a colliding wave 
interpretation. 

It may occur several mixed cases depending on the degree of the polynomials 
appearing in the non-killingian sector, for instance,
\begin{equation}
\frac{dp^2}{P_4}- \frac{dq^2}{Q_2}, \quad \frac{dp^2}{P_3}- \frac{dq^2}{Q_2},
\end{equation}
where the subindice indicates the degree of the polynomial we are dealing 
with;  the possibility of interchanging $P$ by $Q$ and $p$ by $q$ and 
viceversa has to be taken also into account. The case 
$\frac{dp^2}{P_3}- \frac{dq^2}{Q_3}$ can be treated in the manner we did 
with the fourth degree polynomial. 

Let us consider the case of $\frac{dp^2}{P_4}- \frac{dq^2}{Q_2}$. Introducing 
the null variables $( \widetilde u, \widetilde v)$,
\begin{equation}
\left. \begin{array}{l}
d{\widetilde u} \\ 
d{\widetilde v}
\end{array} \right\}  =\frac{dp}{\sqrt{P_4(p)}} \pm  \frac{dq}{\sqrt{Q_2(q)}}  
\label{kat}\, ,
\end{equation}
in the case when $Q_2$ can be brought to the form $1-q^2$, they become
\begin{equation}
\left. \begin{array}{l}
{\widetilde u} \\ 
{\widetilde v}
\end{array} \right\}  = \mu_{\phi} F(\phi, k_{\phi}) \pm \arcsin {q},  
\label{kati}\, ,
\end{equation}
therefore
\begin{eqnarray}
\frac{\widetilde u - \widetilde v}{2}&=& \arcsin{q}, \\
\frac{\widetilde u + \widetilde v}{2}&=& \mu_{\phi} F(\phi, k_{\phi})=
{\mu_\phi} \int^{\phi}
\frac{dx}{\sqrt{1-k_\phi^2 \sin^2 x}}= \mu_{\phi} z, \quad \phi= {\rm am} z,
\end{eqnarray}
hence
\begin{equation}
q= \sin( \frac{\widetilde u - \widetilde v}{2}), \quad p= p
( \frac{\widetilde u + \widetilde v}{2})
\end{equation}
where the relation of $p$ in terms of $\widetilde u, \widetilde v$ depend 
on the roots of the polynomial $P_4$. 
The other case with $P_3, Q_2$ can be treated in similar way.

For completeness we also briefly comment the following cases

\noindent{(i)} 
\begin{equation}
\frac{dp^2}{P_{4,3}(p)} -  \frac{dq^2}{Q_1(q)}  \to 
\frac{dp^2}{P_{4,3}(p)} -  \beta (dq^{ \frac{1}{2}})^2, 
\quad q>0, \beta >0,
\end{equation}
the corresponding $\widetilde u, \widetilde v$ are
\begin{equation}
\left. \begin{array}{l}
{\widetilde u} \\ 
{\widetilde v}
\end{array} \right\}  =\mu_{\phi}F(\phi, k_{\phi}) \pm  \beta q^{\frac{1}{2}},  
\end{equation}
thus
\begin{equation}
p= p( \frac{\widetilde u + \widetilde v}{2}), \quad 
q= ( \frac{\widetilde u - \widetilde v}{2 \beta})^2,
\end{equation}

\noindent{(ii)} 
\begin{equation}
\frac{dp^2}{P_{2}(p)} -  \frac{dq^2}{Q_2(q)}  \to 
\frac{dp^2}{1-p^2} -  \frac{dq^2}{ 1-q^2}, 
\end{equation}
the corresponding $\widetilde u, \widetilde v$ are
\begin{equation}
\left. \begin{array}{l}
{\widetilde u} \\ 
{\widetilde v}
\end{array} \right\}  = \arcsin{p} \pm \arcsin{q},  
\end{equation}
hence
\begin{equation}
p= \sin( \frac{\widetilde u + \widetilde v}{2}), \quad 
q= \sin( \frac{\widetilde u - \widetilde v}{2}),
\end{equation}

\noindent{(iii)} 
\begin{equation}
\frac{dp^2}{P_{4,3}(p)} - {dq^2}, 
\end{equation}
the corresponding $\widetilde u, \widetilde v$ are
\begin{equation}
\left. \begin{array}{l}
{\widetilde u} \\ 
{\widetilde v}
\end{array} \right\}  =  \mu_{\phi}F(\phi, k_{\phi}) \pm {q},  
\end{equation}
hence
\begin{equation}
p= p( \frac{\widetilde u + \widetilde v}{2}), \quad 
q=  \frac{\widetilde u - \widetilde v}{2},
\end{equation}

\noindent{(iv)} 
\begin{equation}
\frac{dp^2}{P_{2}(p)} - \frac{dq^2}{Q_1} \to 
\frac{dp^2}{1-p^2} - \beta (dq^{ \frac{1}{2}})^2, \quad \beta >0, q >0, 
\end{equation}
the corresponding $\widetilde u, \widetilde v$ are
\begin{equation}
\left. \begin{array}{l}
{\widetilde u} \\ 
{\widetilde v}
\end{array} \right\}  =  \arcsin{p} \pm \beta q^{\frac{1}{2}},  
\end{equation}
hence
\begin{equation}
p= \sin( \frac{\widetilde u + \widetilde v}{2}), \quad 
q=  (\frac{\widetilde u - \widetilde v}{2 \beta})^2,
\end{equation}

\noindent{(v)} 
\begin{equation}
\frac{dp^2}{P_{1}(p)} -  \frac{dq^2}{Q_1(q)}  \to 
(dp^{ \frac{1}{2}})^2 -   (dq^{ \frac{1}{2}})^2, 
\quad p>0, q >0,
\end{equation}
the corresponding $\widetilde u, \widetilde v$ are
\begin{equation}
\left. \begin{array}{l}
{\widetilde u} \\ 
{\widetilde v}
\end{array} \right\}  = p^{\frac{1}{2}} \pm  q^{\frac{1}{2}},  
\end{equation}
thus
\begin{equation}
p= ( \frac{\widetilde u + \widetilde v}{2})^2, \quad 
q= ( \frac{\widetilde u - \widetilde v}{2})^2,
\end{equation}

\subsection{Different transformations of $p(\widetilde u, \widetilde v)$ 
and $q(\widetilde u, \widetilde v)$.}
In this subsection, we present all possible cases depending upon the roots
of the polynomials $P(p)$ and $Q(q)$ and the corresponding
transformation.
\begin{enumerate}

\item For $P$ and $Q$ with all real and different roots, i.e.  
\begin{eqnarray}
P&=& \alpha_4  \epsilon_\pm \left[p-p_1\right]\left[p-p_2\right]
\left[p-p_3\right]\left[p-p_4\right], \qquad \epsilon_\pm = \pm 1 , \, 
\alpha_4>0\, ,\\ 
Q&= & \beta_4 \nu_\pm \left[q-q_1\right]\left[q-q_2\right]
\left[q-q_3\right]\left[q-q_4\right], \qquad \nu_\pm = \pm 1, \, \beta_4 >0  
\label{cu}\, ,
\end{eqnarray}
The corresponding transformations read as follows
\vspace{0.5cm}

\begin{tabular}{|c|c|c|c|c|}   \hline\hline
$P_4(p)$ & $\epsilon_{\pm}$  & Interval & $p({\widetilde u},
{\widetilde v})=$ &  $k^2$ \\ \hline\hline 
four &$\epsilon_+ $  & $p_1 \leq  p$ or $p \leq p_4$  &
$\frac{p_1 p_{42} 
- p_2 p_{41}\left({\rm sn} \frac{1}{2\mu_\phi} ({\widetilde u} +{\widetilde v})
\right)^2}{p_{42} - p_{41}\left({\rm sn} \frac{1}{2\mu_\phi} ({\widetilde u} +
{\widetilde v})\right)^2}$ &   \\  \cline{2-4}
real & $\epsilon_+$ & $p_3 \leq p \leq p_2$  &
$\frac{p_3 p_{42} 
- p_4 p_{32}\left({\rm sn} \frac{1}{2\mu_\phi} ({\widetilde u} +{\widetilde v})
\right)^2}{p_{42} - p_{32}\left({\rm sn} \frac{1}{2\mu_\phi} ({\widetilde u} +
{\widetilde v})\right)^2}$  & $(\alpha_1, \alpha_2, \alpha_4, \alpha_3)$ \\ 
\cline{2-5}
roots & $\epsilon_-$ & $p_4  \leq  p \leq p_3$ &
$\frac{p_4 p_{31} 
+ p_1 p_{43}\left({\rm sn} \frac{1}{2\mu_\phi} ({\widetilde u} +{\widetilde v})
\right)^2}{p_{31} + p_{43}\left({\rm sn} \frac{1}{2\mu_\phi} ({\widetilde u} +
{\widetilde v})\right)^2}$ &    \\ \cline{2-4}
 & $\epsilon_-$ & $p_2 \leq  p \leq p_1$  &
$\frac{p_2 p_{31} 
- p_3 p_{21}\left({\rm sn} \frac{1}{2\mu_\phi} ({\widetilde u} +{\widetilde v})
\right)^2}{p_{31} - p_{21}\left({\rm sn} \frac{1}{2\mu_\phi} ({\widetilde u} +
{\widetilde v})\right)^2}$  & $(\alpha_3, \alpha_2, \alpha_4, \alpha_1)$  \\  
\hline
\end{tabular}
\vspace{0.5cm}

In the transformations given above 
we must understand that $ \widetilde u = 2 \arcsin {u}$ and 
$ \widetilde v = 2 \arcsin {v}$, such that 
$\widetilde u \pm \widetilde v = 2 \arcsin(uV \pm vU)$ and 
$\mu_\phi$ 
has to be understood
as $\mu_\phi/\sqrt{\alpha_4}$ when $\alpha_4 \neq 1$.
For $Q(q)$ the same Table is  valid making the changes: 
$p \to q$, $p_i \to q_i$,
$ \epsilon_{\pm} \to \nu_{\pm}$, $\mu_{\phi} \to \mu_{\theta}$ and 
$ \widetilde u + \widetilde v \to \widetilde u - \widetilde v$, 
$s_i \to \sigma_i$ and $t_i \to \tau_i$.
The same applies in the following cases.

\item For the case in which $P$ and $Q$ possess two real and different 
roots, and two  complex roots, i.e.
\begin{eqnarray}
P&=& \alpha_4 \epsilon_\pm \left[p-p_1\right]\left[p-p_2\right]
\left[p-(s_1+it_1)\right]\left[p-(s_1-it_1)\right], \quad \alpha_4 > 0, \\ 
Q&= \beta_4&\nu_\pm \left[q-q_1\right]\left[q-q_2\right]
\left[q-({\widetilde \sigma}_1+i
{\widetilde \tau}_1)\right]\left[
q-({\widetilde \sigma}_1-i{\widetilde \tau}_1)\right], \quad \beta_4 > 0, 
\end{eqnarray}
the transformations are given by
\vspace{0.5cm}

\begin{tabular}{|c|c|c|c|c|}   \hline\hline
$P_4(p)$ & $\epsilon_{\pm}$  & {\it Interval} & $p({\widetilde u},
{\widetilde v})=$  & $k^2$ \\ \hline\hline 
two complex  and & $\epsilon_+ $  & $ - \infty < p < \infty$  &
$ \frac{p_1 +p_2}{2} - \frac{p_1 - p_2}{2}  \left( \frac{ \nu_{\theta} 
- {\rm cn} \frac{1}{2\mu_\phi} ({\widetilde u} +{\widetilde v})}{1 
- \nu_{\theta}{\rm cn} \frac{1}{2\mu_\phi} ({\widetilde u} +{\widetilde v})}
\right)$ & 
$ \left( \sin ( \frac{\theta_1- \theta_2}{2}) \right)^2$  \\  
two real roots & & & & \\ \hline
\end{tabular}
\vspace{0.5cm}

\item  In the case $P$ and $Q$ with four complex roots, respectively, i.e.
\begin{eqnarray}
P&=& \alpha_4 \left[p-(s_1+it_1)\right]\left[p-(s_1-it_1)\right]
\left[p-(s_2+it_2)
\right]\left[p-(s_2-it_2)\right], \\
Q&=& \beta_4 \left[q-({\widetilde \sigma}_1+i{\widetilde \tau}_1)
\right]\left[q-({\widetilde \sigma}_1-i{\widetilde \tau}_1)
\right]\left[q-({\widetilde \sigma}_2+i{\widetilde \tau}_2)
\right]\left[q-({\widetilde \sigma}_2-i{\widetilde \tau}_2)\right],  
\end{eqnarray}
the transformations are now
\vspace{0.5cm}

\begin{tabular}{|c|c|c|c|c|}   \hline\hline
$P_4(p)$ & $\epsilon_{\pm}$  & Interval & 
$p({\widetilde u},{\widetilde v})=$ 
& $k^2$ \\ \hline\hline 
$s_1 > s_2$ & &  &
$s_1+t_1\left[
 \frac{{\rm sn}\frac{1}{2\mu_\phi} 
({\widetilde u} + {\widetilde v}) {\rm cos}\left(\frac{\theta_3}{2}+
\frac{\theta_4}{2} \right) + {\rm cn}\frac{1}{2\mu_\phi} ({\widetilde u} +
{\widetilde v}) {\rm sin} \left(\frac{\theta_3}{2}+\frac{\theta_4}{2} \right) }
{{\rm cn}\frac{1}{2\mu_\phi} 
({\widetilde u} +{\widetilde v}){\rm cos}\left(\frac{\theta_3}{2}+
\frac{\theta_4}{2} \right) - {\rm sn}\frac{1}{2\mu_\phi} ({\widetilde u}+ 
{\widetilde v}) {\rm sin} \left(\frac{\theta_3}{2}+\frac{\theta_4}{2} \right)}
\right]$ & $\sin^2 \theta_5$  \\  
 & $\epsilon_+$ & $- \infty < p < \infty$  & &     \\ \cline {1-1} \cline{4-4}
$s_1=s_2$ &   & &
$s_1-t_1\left[
 \frac{{\rm sn}\frac{1}{2\mu_\phi} 
({\widetilde u} + {\widetilde v}) {\rm cos}\left(\frac{\theta_3}{2}+
\frac{\theta_4}{2} \right) + {\rm cn}\frac{1}{2\mu_\phi} ({\widetilde u} +
{\widetilde v}) {\rm sin} \left(\frac{\theta_3}{2}+\frac{\theta_4}{2} \right) }
{{\rm cn}\frac{1}{2\mu_\phi} 
({\widetilde u} +{\widetilde v}){\rm cos}\left(\frac{\theta_3}{2}+
\frac{\theta_4}{2} \right) - {\rm sn}\frac{1}{2\mu_\phi} ({\widetilde u}+ 
{\widetilde v}) {\rm sin} \left(\frac{\theta_3}{2}+\frac{\theta_4}{2} \right)}
\right]$ &  \\ 
B $t_1>t_2$ &   &  &  &    \\ \hline
\end{tabular}

\end{enumerate}
\vspace{0.5cm}

We include the transformations for the case of polynomials of third degree:
\vspace{0.5cm}

\begin{tabular}{|c|c|c|c|c|}   \hline\hline
$P_3(p)$  & $\epsilon_{\pm}$  & Interval & $p({\widetilde u},{\widetilde v})=$ 
& $k^2$ \\ \hline\hline 
three & $\epsilon_+ $  & $p_3 \leq  p \leq p_2$  &
$\frac{ p_{32}}{p_3}\left({\rm sn} \frac{1}{2\mu_\phi} ({\widetilde u} +
{\widetilde v})\right)^2$ &  $\frac{p_{32}}{p_{31}}$ \\  \cline{3-4}
real &  & $p_1 \leq p $  &
$\frac{p_1 - p_{2}\left({\rm sn} \frac{1}{2\mu_\phi} ({\widetilde u} +
{\widetilde v})\right)^2}{1- \left({\rm sn} \frac{1}{2\mu_\phi} 
({\widetilde u} +
{\widetilde v})\right)^2 }$
&   \\ \cline{2-5}
roots & $\epsilon_-$ & $p \leq  p_1$ &
$\frac{ p_1 \left({\rm sn} \frac{1}{2\mu_\phi} ({\widetilde u} +
{\widetilde v})\right)^2 - p_{31}}{\left({\rm sn} \frac{1}{2\mu_\phi} 
({\widetilde u} +
{\widetilde v})\right)^2}$ & $ \frac{p_{21}}{p_{31}}$   \\ \cline{3-4}
 & & $p_2 \leq  p \leq p_1$  &
$\frac{p_2 p_{31} -p_3 p_{21}\left({\rm sn} \frac{1}{2\mu_\phi} 
({\widetilde u} +
{\widetilde v})\right)^2 }{p_{31}- p_{21}\left({\rm sn} \frac{1}{2\mu_\phi} 
({\widetilde u} +
{\widetilde v})\right)^2}$ &  \\  \hline\hline
two & $\epsilon_+$ & $p_1 \leq p$ & $p_1 - \frac{c_1 \left( 1- {\rm cn}
\frac{1}{2\mu_\phi} ({\widetilde u}+ {\widetilde v})\right)}{\cos \theta_1  
\left( 1- {\rm cn}\frac{1}{2\mu_\phi} ({\widetilde u}
+ {\widetilde v})\right)}$  & \\ \cline{2-3}
complex and & & & & $\left( \sin \left[ \frac{\theta_1}{2} + \frac{\pi}{4}
\right] 
\right)^2$ \\
one real & $\epsilon_{-}$ & $p \leq p_1$ & & \\ \hline
\end{tabular}
\vspace{0.5cm}

For the polynomial $Q$ the same Tables are  valid making the changes: 
$p \to q$, 
$p_i \to q_i$,$\epsilon_{\pm} \to \nu_{\pm}$, $\mu_{\phi} \to \mu_{\theta}$ and 
$ \widetilde u + \widetilde v \to \widetilde u - \widetilde v$, 
$s_i \to \sigma_i$ and $t_i \to \tau_i$.

All possible combinations between the 
transformations corresponding to the different possible ranges of $p$ and $q$  
should be taken into account. Moreover, it is straightforward to perform such 
combinations by using the explicit transformations in 
tables given above.

\subsection{Limiting transition}

For completeness we include the limiting procedure to obtain from the fourth 
order polynomials the third and second degree ones.

\begin{eqnarray}
P&=& \epsilon_{\pm} a_4 (p-p_1)(p-p_2)(p-p_3)(p-p_4) \nonumber  \\ 
&=& a_4 p_1(\frac{p}{p_1}-1)(p-p_2)(p-p_3)(p-p_4)  \nonumber  \\  
&=& \alpha_4(\frac{p}{p_1}-1)(p-p_2)(p-p_3)(p-p_4),
\end{eqnarray}
Taking the limit when $p_1 \to \infty$ we recover a third degree polynomial:
\begin{equation}
\lim_{p_1 \to \infty} P_4 = \epsilon_{\pm} \alpha_4 (p-p_2)(p-p_3)(p-p_4) 
= \epsilon _{\pm} \alpha_4 P_3,
\end{equation}

Analogously we take the limit on $P_3$ when $p_2 \to \infty$
to get a second order polynomial:

\begin{eqnarray}
\lim_{p_2 \to \infty} P_3 &=&  \lim_{p_2 \to \infty} 
\left\{ \epsilon_{\pm} a_4 p_2 (\frac{p}{p_2}-1)(p-p_3)(p-p_4) \right\} 
\nonumber \\
&=& \lim_{p_2 \to \infty} \left\{ \epsilon_{\pm} \alpha_3 
(\frac{p}{p_2}-1)(p-p_3)(p-p_4) \right\} = \epsilon_{\pm} 
\alpha_3 (p-p_3)(p-p_4)
= \epsilon_{\pm} \alpha_3 P_2,
\end{eqnarray}

The corresponding procedure can be achieved for $Q_4$ to obtain a 
second order polynomial.

\subsection{Concrete example}
As a concrete example let us consider, for instance, the case $P$ with four 
real and different roots and $Q$ also with four real and different roots, in 
which the coefficients of the higher degree are 
$1$, and the roots of $P$ and $Q$ are denoted by 
$p_i$ and $q_i$ respectively. For the elliptic integral 
depending on the $p$--coordinate, with $P(p)= (p-p_1)(p-p_2)(p-p_3)(p-p_4)$, 
($\epsilon_{ \pm}=1$), one has the relation
\begin{equation}
\int \frac{dx}{\sqrt{P(p)}} = \mu_\phi \int 
\frac{d\phi}{\sqrt{1-k_\phi^2 
\sin^2 \phi}} = \mu_\phi F(\phi,k)\, , 
\end{equation}  
where the explicit relation between $p$ and $\phi$ reads
\begin{eqnarray}
p&=& \frac{p_1 p_{42} - p_2 p_{41} \sin^2 \phi}{p_{42} - p_{41} \sin^2 \phi}\,
 , \nonumber \\
\sin^2 \phi &=& \frac{p_{42}}{p_{41}} \left( \frac{p-p_1}{p-p_2}\ \right)\, , 
\quad \rightarrow
\quad \phi = \arcsin \left(\pm \sqrt{\frac{p_{42}}{p_{41}} \left( \frac{p-p_1}
{p-p_2} \right)}\right)
\label{aurrera}\, . 
\end{eqnarray} 
The parameter $k_\phi$, $0<k_\phi^2<1$, is given by
\begin{equation}
k^2= (p_1, p_2, p_4, p_3) = \left( \frac{p_1-p_4}{p_1-p_3} \right) 
\left( \frac{p_2-p_3}{p_2-p_4} \right) \, , 
\quad {\rm and} \quad \mu = \frac{\mu_{\phi}}{\sqrt{\alpha_4}} = \frac{2}{
\sqrt{p_{31}p_{42}}}\, .
\end{equation}
It is straightforward to find the analogous expression for the elliptic
integral depending on the $q$--coordinate.  The explicit expression of the
coordinates $p$ and $q$ through ${\widetilde u}$ and ${\widetilde v}$ is
the following
\begin{eqnarray}
p({\widetilde u},{\widetilde v})&=& \frac{p_1 p_{42} 
- p_2 p_{41}\left({\rm sn} \frac{1}{2\mu_\phi} ({\widetilde u} +{\widetilde v})
\right)^2}{p_{42} - p_{41}\left({\rm sn} \frac{1}{2\mu_\phi} ({\widetilde u} +
{\widetilde v})\right)^2} \label{x1}\, , \\
q({\widetilde u},{\widetilde v})&=& \frac{q_1 q_{42} - q_2 
q_{41}\left({\rm sn} \frac{1}{2\mu_\theta} ({\widetilde u} 
- {\widetilde v})\right)^2}{q_{42} - q_{41} 
\left({\rm sn} \frac{1}{2\mu_\theta}({\widetilde u} - {\widetilde v})
\right)^2} 
\label{ongi}\, .
\end{eqnarray}

In terms of the null variables $u$ and $v$ which bring the line element to 
the form
\begin{equation}
\frac{dp^2}{P_4}- \frac{dq^2}{Q_4} = d \widetilde u d \widetilde v = 
4 \frac{dudv}{UV}
\end{equation}
the $p$ and $q$ variables amount to 
\begin{eqnarray}
p( u, v)&=& \frac{p_1 p_{42} 
- p_2 p_{41}\left({\rm sn} \frac{1}{\mu_\phi} ( \arcsin(uV + vU))
\right)^2}{p_{42} - p_{41}\left({\rm sn} \frac{1}{2\mu_\phi} 
( \arcsin(uV + vU))\right)^2} \label{x10}\, , \\
q( u, v)&=& \frac{q_1 q_{42} - q_2 
q_{41}\left({\rm sn} \frac{1}{2\mu_\theta} ( \arcsin(uV - vU))\right)^2}
{q_{42} - q_{41} 
\left({\rm sn} \frac{1}{2\mu_\theta}( \arcsin(uV - vU))
\right)^2}
\label{x12}\, ,
\end{eqnarray}

On the other hand, the relations between $\phi$ and $\theta$ with $p$ and 
$q$ are  
\begin{equation}
\phi \pm \theta= \arcsin \sqrt{\frac{p_{42}}{p_{41}} 
\left( \frac{p-p_1}{p-p_2}\right)}  \pm
\arcsin \sqrt{\frac{q_{42}}{q_{41}} \left( \frac{q-q_1}{q-q_2}\right)}
\, ,
\end{equation}
where $\phi$ and $\theta$ stand correspondingly for the arguments of the
elliptical Legendre integrals related with the integration of $P$ and
$Q$. We define the auxiliary non-null variables $u'$ and $v'$ as
\begin{equation}
u' = \sin \left( \frac{\phi + \theta}{2} \right) \, , \quad  v' = \sin 
\left( \frac{\phi - \theta}{2} \right)\, , 
\end{equation}  
with
\begin{equation}
\left. \begin{array}{l}
\phi \cr 
\theta
\end{array} \right\}= \arcsin u' \pm \arcsin v'=\arcsin \left[u' \sqrt{1-v'^2}  
\pm v' \sqrt{1-u'^2} 
\right]= \arcsin \left[u'V' \pm v'U' \right]
\label{fi}\, ,  
\end{equation}
where 
\begin{equation}
V'=\sqrt{1-v'^2}\, , \quad  {\rm and}\quad  U'=\sqrt{1-u'^2}
\label{teta}\, .
\end{equation}
Thus, when we substitute (\ref{fi}) and (\ref{teta}) into (\ref{x1}) and
(\ref{ongi}) it is staightforward to find
\begin{eqnarray}
p(u',v')&=& \frac{p_1 p_{42} - p_2 p_{41}\left(u'V'+v'U'\right)^2}{p_{42} - 
p_{41}
\left(u'V'+v'U'\right)^2} \label{p}\, , \\
q(u',v')&=& \frac{q_1 q_{42} - q_2 
q_{41}\left(u'V'-v'U'\right)^2}
{q_{42} - q_{41} \left(u'V'-v'U'\right)^2} 
\label{q}\, .
\end{eqnarray}
Moreover, from (\ref{beto1}) and (\ref{fi}), the advanced and retarded time 
coordinates ${\widetilde u}$ and
${\widetilde v}$ are related with the variables $u'$ and $v'$ through
\begin{eqnarray}
\left. \begin{array}{l}
d{\widetilde u}\cr 
d{\widetilde v}
\end{array} \right\}&= &\left[\frac{\mu_\phi}{\sqrt{1-k_\phi^2 (u'V'+vU')^2}} 
\pm  \frac{\mu_\theta}
{\sqrt{1-k_\theta^2 (u'V'-v'U')^2}}\right] \frac{du'}{U'} \nonumber \\
&+&\left[\frac{\mu_\phi}{\sqrt{1-k_\phi^2 (u'V'+v'U')^2}} \mp  \frac{\mu_\theta}
{\sqrt{1-k_\theta^2 (u'V'-v'U')^2}}\right] \frac{dv'}{V'} 
\label{null} \, .
\end{eqnarray} 

According to our general procedure (\ref{null}) can be inverted, yielding 
$u'=u'({\widetilde u},{\widetilde v})$ and $v'=v'({\widetilde u},
{\widetilde v})$. 
 
In the limit of polynomials of second degree , i.e. $P_2$, $Q_2$, one has 
$k_\phi=k_\theta=0$, and $\mu_\phi, \mu_\theta \rightarrow 1$, hence one 
recovers from (\ref{null}) the widely used retarded and advanced time 
coordinates
\begin{equation}
d{\widetilde u}=2\frac{du'}{U'}\, \quad d{\widetilde v}=2\frac{dv'}{V'}
\label{null1}\, ,
\end{equation}  
In the next Section we present an explicit class of colliding wave solutions. 
\section{A class of colliding waves}

Let us consider an explicit solution of the Einstein-Maxwell equations 
in region I, given by the following coframe in coordinates $(x,p,q,y)$:
\begin{eqnarray}
\vartheta^{\hat{0}}&=&\frac{1}{H}\sqrt{\frac{\Delta}{Q}}\;dq, \quad
\vartheta^{\hat{1}}=\frac{1}{H}\sqrt{\frac{Q}{\Delta}}\; 
(dx+p^2dy)\, , \nonumber \\ 
\vartheta^{\hat{2}}&=&\frac{1}{H}\sqrt{\frac{\Delta}{P}}\;dp, 
\quad 
\vartheta^{\hat{3}}=\frac{1}{H}\sqrt{\frac{Q}{\Delta}}\; (dx-q^2dy)
\label{coframe} \, .
\end{eqnarray} 
Here we have the functions
$H=H(p,q)$, $P=P(p)$, $Q=Q(q)$, and
$\Delta=\Delta(p,q)$. The coframe is assumed to be orthonormal 
\begin{equation}
g=o_{\alpha\beta}\,\vartheta^\alpha\otimes\vartheta^b\,.
\end{equation} 
Then the metric explicitly reads 
\begin{equation} 
g=\frac{1}{H^2} \left \{
\frac{Q}{\Delta} \left( dx+ p^2 \, dy\right)^2
- \frac{\Delta}{Q}\, dq^2 + \frac{\Delta}{P}
\,dp^2 + \frac{P}{\Delta} \left( dx -q^2 dy
\right)^2 \right \}
\label{ortho} \, .  
\end{equation}
with
\begin{eqnarray} 
H(p,q) &:=& 1 - \mu p q \, , \nonumber\\ 
P(p) &:=& b-g^2 + 2 n p -
\varepsilon p^2 +2 m \mu \, p^3 + \left( -\frac{\lambda}{3}- \mu^2 b -
\mu^2 e^2 \right)\, p^4 \, , \nonumber\\ 
Q(q) &:=&-(b+ e^2) + 2 m q - \varepsilon q^2 + 2 n \mu \, q^3 
+ \left( \frac{\lambda}{3} + \mu^2 b -
\mu^2 g^2 \right)\, q^4 \, , \nonumber\\ 
\Delta(p,q)&:=& p^2 + q^2 \, ,
\label{sol1}
\end{eqnarray}
while the nonvanishing electromagnetic field components are given by
\begin{eqnarray} 
F_{xq} &:=& \frac{(p^2-q^2)e-2gpq}{2 \tilde \Delta^2} \, , \nonumber\\ 
F_{yq} &:=& p^2 F_{xq} \, , \nonumber\\ 
F_{xp} &:=&  \frac{g(q^2-p^2)-2epq}{\tilde \Delta^2} \, , \nonumber\\ 
F_{yp}&:=& - q^2 F_{xp} \, .
\label{solutions}
\end{eqnarray}
In order to express our solutions (\ref{sol1}) and (\ref{solutions}) in
terms of the advanced and retarded time coordinates ${\widetilde u}$,
${\widetilde v}$, one has to use the inversion procedure,presented in 
the previous section, yielding 
$p=p({ u},{v})$ and $q=q({ u},{ v})$.
The explicit representation requires the use of 
tables for the ${\rm sn}$ function.

Our class of solutions, defined in region I, can be extended to the full 
spacetime by introducing the Heaviside step function 
\begin{equation}
\Theta ({x})= \cases{1\, , \quad {x} \ge 0\cr
                  0\, , \quad {x}<0}\; ,
\end{equation} 
with $\Theta^2({x}) = \Theta({x})$, 
and performing the following transformation 
\begin{equation}
u \rightarrow \Theta( u) u\, , \qquad {\rm and} \qquad 
v \rightarrow \Theta(v) v
\label{ttt}\, .
\end{equation}
For the case of real roots of the polynomials, treated 
extensively in Sec. III.B, we obtain for region II 
\begin{eqnarray}
p({ u},{ v})&=& \frac{p_1 p_{42} 
- p_2 p_{41}\left({\rm sn} \frac{1}{2\mu_\phi} {\widetilde v}
\right)^2}{p_{42} - p_{41}\left({\rm sn} \frac{1}{2\mu_\phi} 
{\widetilde v}\right)^2} \label{x11}\, , \\
q({ u},{ v})&=& \frac{q_1 q_{42} - q_2 
q_{41}\left({\rm sn} \frac{1}{2\mu_\theta}{\widetilde v}\right)^2}
{q_{42} - q_{41} \left({\rm sn} \frac{1}{2\mu_\theta}
{\widetilde v}\right)^2} 
\label{y11}\, ,
\end{eqnarray}
where
\begin{equation}
{\rm sn} \frac{\widetilde v}{2\mu} = {\rm sn} \left[\frac{1}{\mu}
\arcsin v \right]\, .
\end{equation}
For region III
\begin{eqnarray}
p({ u},{ v})&=& \frac{p_1 p_{42} 
- p_2 p_{41}\left({\rm sn} \frac{1}{2\mu_\phi} {\widetilde u}\right)^2}
{p_{42} - p_{41}\left({\rm sn} \frac{1}{2\mu_\phi} 
{\widetilde u}\right)^2} \label{x111}\, , \\
q({u},{ v})&=& \frac{q_1 q_{42} - q_2 
q_{41}\left({\rm sn} \frac{1}{2\mu_\theta}{\widetilde u}\right)^2}
{q_{42} - q_{41} 
\left({\rm sn} \frac{1}{2\mu_\theta}{\widetilde u}\right)^2} 
\label{y111}\, ,
\end{eqnarray}
where
\begin{equation}
{\rm sn} \frac{\widetilde u}{2\mu} = {\rm sn} \left[\frac{1}{\mu}
\arcsin u \right]\, .
\end{equation}
 
At this point it is important to mention that not every cylindrically 
symmetric spacetime, even in vacuum, satifies the requirements of Ernst 
\cite{ernst} for being interpreted as colliding waves, 
only certain classes of cylindrically symmetric solutions can 
be thought of as solutions generated by collision of waves. 

Moreover, up to here our analysis has been concerned mostly with metrical 
aspects leaving aside the task of having suitable field and matter 
energy--momentum tensors. As it is well known, each participating wave in 
the head--on collision is assumed to be plane fronted, i.e. it is 
characterized by a covariantly constant null 
eigenvector {\bf $k$}, i.e., $k_{\mu; \nu}=0$, of the Weyl Petrov type N 
conformal tensor. This condition implies a null energy--momentum 
tensor for the electromagnetic field, i.e. a radiation field 
or electrovacuum, ---let us call it null field for short--- into the Einstein 
field equations, c.f. Kramer et. al. \cite{Kram}, Section \S 21.5.  
Therefore, the existence of colliding wave solutions with non--vanishing 
cosmological $\lambda$ term is thus forbidden.

Usually one derives colliding wave solutions using the backward way 
procedure, namely, one starts with a cylindrically symmetric solution  
and an energy--momentum tensor in terms of fields depending on 
the usual two null variables, 
say $u$ and $v$, restricted to a closed interaction 
region, region I in our approach. Furthermore, by accomplishing 
on it the transformations (\ref{ttt}) with Heaviside functions 
one obtains the spacetimes prior to the collision. 
These spacetimes occur to be type N solutions to the Einstein equations in 
vacuum or coupled with null fields, regions II and III in our approach,
where the gravitational and null fields depend solely on a single  variable 
($u$ or $v$). At this level, the colliding wave condition allows only for 
fields which can be matched to the flat Minkowski spacetime (region IV) 
background where the wavefronts propagate 
\footnote{We appreciate the referee's comment which draw our 
attention to this respect.}. In other words, if the cylindrically symmetric 
spacetime is endowed with a cosmological $\lambda$ term, 
one has to set it equal to zero in order to connect smoothly the different 
regions of the collision and in order to avoid meaningless jumps 
of the cosmological constant across the null lines. 
 
On the other hand, it is worthwhile to mention the results by Chandrasekhar 
and Xanthopoulos \cite{chandra3} concerning colliding wave solutions with a 
perfect fluid (stiff matter, fluid pressure $=$ fluid energy density). 
They found that after the collision takes place, the incoming plane waves 
supporting radiative fields, give rise to solutions with stiff matter 
perfect fluid tensor. It is an example of the transformation of massless 
particles describing null trajectories into a perfect fluid in which the 
stream lines describe timelike trajectories.
 
In full, in order to interpret our class of cylindrically symmetric solutions
(\ref{sol1}) as a colliding waves class of solutions, it is compulsory to 
set $\lambda=0$ on it \cite{Kram}.

\section{Discussion}

In this paper we present the formulation of the most 
general class nowadays of colliding waves with fourth degree polynomials 
determining the non--killingian sector, by means of a generalization of the 
advanced and retarded coordinates concept.
As it has been pointed out, after setting the cosmological $\lambda$ term 
equal to zero, this class of solutions describes the
scattering of two noncolinear polarized gravitation plane waves,
having a null electromagnetic field as a source. At the leading edge of 
each colliding type--N gravitational wave, the curvature tensor exhibits 
delta and jump discontinuities.
The former is interpreted as a gravitational impulsive
wave, whereas the latter is attributed to a gravitational shock wave.

As stated above, this class of solutions, defined again in region I, can also 
be extended to the full spacetime by introducing the Heaviside step function 
$\Theta$. 

The electromagnetic field present delta singularities and jump 
discontinuities. However, the Bianchi identities hold in a distributional
sense, see \cite{taub}. 

There are no problems on the right--hand side because the delta type 
singularities of the curvature are multiplied by the smooth distributions 
$\sqrt{1 - \Theta(u) u^2}$ and $\sqrt{1 - \Theta(v) v^2}$, respectively.

On the other hand, with respect to the singularities of the solution, we have 
in Region I, the solution is of type D in the classification of Petrov, 
this is so because the only novanishing Weyl scalar is $\Psi_2$, and 
explicitly, it is given by
\begin{eqnarray}
\Psi_2&=& \frac{H^3}{2 \Delta^3} [2(e^2+g^2)(1+ \mu pq)(p-iq)^2 
\nonumber \\ 
& & +q(q^2-3p^2)(a_3-ia_1)-p(p^2-3q^2)(a_1+ia_3)]
\label{weylsc}\, ,
\end{eqnarray}
where the parameters $a_1=2n, a_3=2m$. The parameters $\epsilon$ and $b$ 
do not appear in the curvature scalar, however the properties of 
the solution depend on them, as we shall see.
The complex structure of $\Psi_2$ is due to the noncolinear 
polarization of the incoming waves (nondiagonal metric, see \cite{Al}).
Since we have the explicit expression for the Weyl scalar, 
Eq. (\ref{weylsc}), we can 
study the behavior of the invariants of the spacetime and determine if the 
solution develops a curvature singularity a finite time after the instant of 
collision. Penrose attributed the development of the singularity in most 
colliding wave solutions to the self-focusing of the gravitational waves on 
scattering. In this case the invariants are given by $I=3 \Psi_2^2$ and 
$J=\Psi_2^3$. Then the singularities of $\Psi_2$
are clearly the same for the invariants.
A singularity occurs when $\tilde \Delta=p^2+q^2=0$, i. e. when simultaneously
$p=0$ and $q=0$. The interesting question is if there is any case in which 
the nature of the roots do not permit that simultaneously $p=0, q=0$. If this 
is the case, then the solution is free of the focusing singularity. A brief 
analysis shows that this occurrence can be avoided
for certain values of the parameters of the solution.

Let us consider the expressions given in Sec III.C for the case of $P$ and $Q$ 
both with four real roots. In this case (\ref{x10}) and (\ref{x12}) express 
$p$ and $q$ in terms of $(u,v)$. Analyzing those expressions 
we see that $p=0$, $q=0$ simultaneously if, for instance,  
$p_{42}=0, p_{41}=0$ and 
$q_{42}=0$, $q_{41}=0$, i. e. when at least three of the four roots coincide: 
$p_4=p_2 = p_1$ and also $q_4=q_2 = q_1$.
The assumption of the occurrence of three equal roots for $P$ and for $Q$
leads to algebraic conditions that must be satisfied by the parameters of the 
solution, $b, g, n, m, \varepsilon, e$ and  $\mu$.
Some of these conditions are incompatible. For instance, in the case 
considered, the  following condition must be valid: 
$ \mu^2 (b+e^2) =\mu^2 (b-g^2)$.
This is only true in the case if $e^2+g^2 = 0$; i. e.  if $e^2+g^2 \ne  0$ 
we do not have that $p=0$ and $q=0$ simultaneously and then the singularity 
fails to occur. 
Then we conclude that in some cases the existence of the electromagnetic field
permits to tune the parameters in 
such a way to avoid the focusing singularity.
A related result was obtained already by Chandrasekhar for a colliding wave
solution with a scalar field \cite{chandra6} and by Garc\'{\i}a et al. in
the framework of metric--affine gauge theories of gravity \cite{glmms,ghms}.


\acknowledgments

We thank Friedrich W. Hehl and Jos\'e Socorro for useful discussions and 
literature hints. We also thank the unknown referee for his valuable remarks.
This research was supported by  CONACyT (M\'exico), grants No. 3544--E9311,
No. 26329E, No. 3692P--E9607, and by the joint German--Mexican project  
DLR--Conacyt E130--2924 and MXI.6.B0a.6A; and also by FONDECyT 
(Chile)--1980891

\end{document}